\def\ltsima{$\; \buildrel < \over \sim \;$}
\def\simlt{\lower.5ex\hbox{\ltsima}}
\def\gtsima{$\; \buildrel > \over \sim \;$}
\def\simgt{\lower.5ex\hbox{\gtsima}}
\documentstyle[11pt,psfig,newpasp,twoside]{article}
\def\edcomment#1{\iffalse\marginpar{\raggedright\sl#1\/}\else\relax\fi}
\reversemarginpar

\begin{document}
\title{Physical Processes at Cosmic Dawn} 
\author{Andrea Ferrara}
\affil{Osservatorio Astrofisico di Arcetri, Firenze, Italy}

\begin{abstract}
In this summary paper I will focus on the four main topics discussed at
this Conference: (i) First luminous sources in the universe, (ii) Reionization, 
(iii) Intergalactic medium and (iv) Galaxy formation. In addition to provide a 
necessarily sketchy summary of the various talks on these subjects, I will also 
outline some of potentially important issues that still need to be fully 
addressed.   
\end{abstract}

\section{Introduction}

The scientific core of this conference has undoubtedly been the physics of the 
interplay between galaxy formation and the intergalactic medium at relatively 
high redshifts, typically above 5. These epochs can still be compared to the
Old Wild West territories which only brave pioneers have dared to explore. 
The goldy mines containing precious information about the dawn of the universe,
when the first luminous sources brightened up and their light unveiled the
already ongoing formation of large numbers of pregalactic systems, are still
not at reach of the yet most powerful experimental devices currently available 
to us.  This situation is going to change soon, but at the moment 
theorists' predictions remain in the realm of sophisticated (and intellectually exciting) speculations. For how uncertain all this might be considered, 
there is a great deal of knowledge that can be gathered even under these 
unfavorable conditions, as witnessed by the impressive momentum gained by the
field in the very recent years. Thus, as observations are slowly, but steadily
filling the gap with theory, this conference has very suitably served the scope
of setting a  common framework among different, nevertheless connected areas in
cosmology: (i) First luminous sources in the universe, (ii) Reionization, 
(iii) Intergalactic medium and (iv) Galaxy formation. In the following I summarize the main results presented and a few thoughts on future directions, open 
problems and ways to attack them.

\section{First Sources}

The first stars in the universe formed at redshifts of order of $\approx 30$. 
In spite of this generally accepted fact, many questions still remain.

\subsection{How Massive where the First Stars ?}

Star formation in the early universe is usually assumed to be understood
more easily because several complicating effects can be neglected to 
a first approximation: among these are magnetic fields, dust grains and 
metal enrichment. Yet little consensus has been reached among various groups.   
It is clear that gravitational collapse induces fragmentation of the first pregalactic 
objects with initial baryonic mass $\approx 10^5 M_\odot$ into smaller clumps      
of typical mass of about $10^{2-3} M_\odot$, which corresponds to the Jeans mass
set by molecular hydrogen cooling. The occurrence of such fragmentation has been shown 
to require the following conditions $\alpha\equiv$(thermal/gravitational)$<0.3$ and $\beta\equiv$(rotational/
gravitational)$<0.3$ on the relevant energy parameters of the collapsing clouds, at 
least for the quasi-isothermal case ($\gamma=1.1$) derived from 1D simulations.     
Subsequent evolution, in some cases studied with the inclusion of cooling lines 
radiative transfer, shows that a quasi-hydrostatic central core is formed whose specific
mass is somewhat dependent on the details of the problem, but typically in the
range $10^{-3}-1 M_\odot$. However, the precise value of the mass might be irrelevant as 
the accretion rate of the infalling gas is found to be rather high, around $10^{-2} 
M_\odot$~yr$^{-1}$, which implies that, in the absence of any effect quenching accretion,
a large fraction of the initial object can become part of the protostar. Pushing this conclusion a bit further, 
one might predict a top heavy IMF for this first generation of sources.  

If the infall can instead be stopped, different scenarios can be envisaged. Standard ways proposed 
to stop the infall do not seem to work under primordial conditions: radiation pressure might be
opacity limited; bipolar flows need some MDH acceleration process and therefore seem to be excluded.
Angular momentum barrier related to disk formation and competitive accretion by companions are      
slightly more promising, but much more difficult to model accurately. 
In addition to these, there are a number of other effects that will deserve more extensive treatment. 
Among these are: i) {\it the role of HD molecules}. If the abundance of this molecule is higher than the value currently 
used by most studies less massive clumps are expected. However, this is at odd with current determinations
of the baryon density parameter from CMB experiments as BOOMERANG; ii) {\it the role of enrichment}. What is the 
value of the critical metallicity above which metal cooling becomes important both in terms of cooling time
and of contribution to the cloud continuum opacity ? iii) {\it the role of FUV radiation}. It is typically found
that H$_2$ is photodissociated in the collapsing cloud if a flux\footnote{$J_{21}=J/10^{-21}$
erg~s$^{-1}$~cm$^{-2}$~Hz$^{-1}$~sr$^{-1}$}  above $J_{21}>10^5$ can be built up before the onset of nuclear reactions in the core.
This can be seen as a viable mechanism for regulating star formation in primordial objects; iv) indications 
for {\it angular momentum transfer} in the core have been brought as possibly due to turbulence. Assessing in more
detail the relevant physics would certainly constitute a considerable progress. Finally, v) {\it interactions between
clumps} in the same halo, strongly depending on the synchronicity of their formation, is an important aspect for
the global evolution of the object and of the possible feedback effects involved.

However, not all is clear even to a first approximation. For example, studies of cylindrical clouds
(as opposed to the more commonly studied cases of spheroidal clouds) find that such configuration leads
to the formation of fragments with separated mass scales: $\approx 200 M_\odot$ for clumps that formed
during the optically thin phase and $\approx 1 M_\odot$ for clumps formed close to the transition to the
optically thick case (essentially the Jeans masses at $n=n_{crit}$ and $n_{\tau=1}$). If formed stars are
to reflect this distribution, a bimodal IMF is expected. Questions remain on the sensitivity of the results 
to the definition of initial conditions and on the degree of bimodality, but at face value this result might 
imply a low number of TypeII SNe per unit stellar mass formed, which in turn might jeopardize searches for 
high-$z$ SNe. The evolutionary properties of such very massive stars start to be derived with some confidence:
stars larger than $M=120 M_\odot$ disappear in a tremendous explosion releasing about 25 times more energy than
their low mass SN counterparts, leaving no compact remnant. However, the large [Co/Fe] ratios observed in halo
stars seem to rule out the fact that the dominant formation mode in the universe was due to stellar masses
larger than 150 $M_\odot$, although considerable uncertainties remain on these results.

\subsection{Feedback Effects on H$_2$}             

FUV radiation, as already mentioned above, might be an efficient self-regulatory mechanism of primordial star formation.
As rule of thumb, photoionizing radiation will tend to increase the molecular hydrogen abundance of a gas cloud due to 
the larger number of electrons available for the H$^{-}$ production channel; Lyman-Werner radiation will act in 
the opposite way because of the induced two-step (Solomon) process dissociation. These two mechanisms are usually jointly  
addressed as {\it radiative feedback}.
Central to the problem of early star formation is to assess to which extent radiation may alter the collapse evolution.
Studies not including the gravitational field of the dark matter halo have come to the conclusion that the cooling mass
above which the collapse is insensitive to radiative feedback effects, due to self-shielding, is of the order
of $4\times 10^6 J_{21}^{-1/3} M_\odot$. However detailed, this results does not allow to appreciate the dependence on the
virialization redshift of the object for which a proper cosmological context must be defined. A couple of groups have 
obtained result on this issue, both for $\Lambda$CDM and CDM models. The conclusion is that the threshold circular velocity
for collapse is around 30 km~s$^{-1}$, insensitive to the details of radiation transfer. Ideally, a more global study coupling 
radiative transfer and hydrodynamics to yield the dependence of the self-shielding mass on $J_{21}$ at various epochs
would be the final aim. This study is particularly important also in view of the fact that the radiation flux is 
probably dominated by proximity effects of sources at $z>15$ and by a more uniform background at later epochs.   

\section{Reionization}                          

There are at least a few main features on which consensus is reached on how reionization of the intergalactic
medium proceeded. First, individual H$_{\rm II}$ regions developed around the ionizing sources. In the second phase,
these ionized regions grow in number and size until they overlapped, producing a sudden increase of the photon mean free
path. Finally, as the underdense regions (voids) were completely ionized, the ionization fronts started to 
penetrate into overdense regions (clumps and filaments), thus bringing the reionization process to completion.
These general features constitute the minimal common model accepted by all the groups involved in this research area. 	
      
Before we discuss more specific results, it is instructive to describe  the type of sources that have 
been considered up to now in reionization calculations. By far the most popular models assume stellar type objects,
typically PopIIIs and mini-galaxies with Salpeter IMF (with exceptions represented by more top-heavy 
IMFs following the new proposal by Larson). Radiative feedback of the type mentioned above and/or {\it stellar feedback},
i.e. due to SN energy deposition, have been included only by a very limited number of studies, the same statement 
holding for the consideration of Spectral Energy Distributions of the sources more consistent with their (extremely) 
low or zero metallicity. Reionization by QSOs has been much less explored, partly because of technical problems of 
simulations in dealing with the large dynamic range required, and also because of the uncertainties on 
the contribution (and the existence!) of accretion-fueled    radiation beyond redshifts of 6. A third proposal involves
production  of X-ray radiation from Compton up-scattering of CMB photons by electron accelerated at the shock produced by the explosion of the first SNe. This last model would predict sensibly different features of the reionization (for example 
in terms of CMB secondary anisotropies) and therefore should be readily testable as soon as the necessary data 
become available.      

With this background we can now turn to the discussion of some general, but often debated, results.
The epoch of reionization seems to have occurred some time in the redshift range $7 < z_i < 15$:
although the precise value of $z_i$ depends on the parameters/cosmology the range appears to be 
rather robust. Typically, 5-10 ionizing photons per baryon are available from the ionizing sources, and
the reionization can be completed by forming no more than a few percent of the stars that we see today. 
The intensity of the UV background jumps by a factor 10 to 100 (according to the results presented) 
at overlap, reaching a value of $J_{21}=0.1$ afterwards. Associated with the inhomogeneous ionization field 
produced by a non uniform distribution of sources and IGM density are the CMB secondary anisotropies whose 
angular power spectrum peaks at sub-arcminute scales with amplitude $\Delta T/T \approx 10^{-6}$. 
Magnetic field seeds are produced via the Biermann battery as the ionization fronts blow out of the 
ISM of the galaxies and during the final ionization of overdense regions, reaching magnitudes of order
$10^{-19}$~G before redshift of 10, a values high enough to become of interest for subsequent turbulent/dynamo 
amplification.  Finally, after reionization both the Jeans and the filtering (i.e. the smoothing length
of perturbations on small scales due to finite gas pressure) scales increase abruptly by several orders of magnitude.

As usually, large uncertainties remain  and what follows is only an incomplete list. First, we would like to
determine the baryon-to-star conversion factor in early galaxies and if this quantity is a function of the 
object mass. For example, it is possible that this factor is much smaller for halos cooling via the 
rather unefficient molecular cooling than for larger, hydrogen line cooling dominated protogalaxies. 
The second uncertain parameter is the escape fraction of ionizing photons from the overdense regions in which
they are produced. Current detailed models generally put upper limits of about 10\% on $f_{esc}$, and are in
agreement with observations of the local universe. The situation might be different for low mass objects, in 
which a few OB stars may ionize the entire ISM of the galaxy in a time scale shorter than the lifetime of the
stars themselves. A related question concerns the IGM clumping factor (some ambiguity indeed exists in the 
definition of this quantity and the  escape fraction above): as  some halos will be too small or sensitive to 
radiative feedback to collapse, they will float around as dark  objects (or {\it sterile} halos) until they are 
photoionized. Their relative high density will pose stronger requirements on the photon-to-baryon budget ratio
required to reionize the IGM.         

\subsection{Observational Probes of the End of Dark Ages}

A remarkable number of possible observational tests of the epochs close to $z_i$ have been proposed and
to discuss them all in detail would be outside the purpose of the present review. Therefore, in the following 
I just list the proposed experiments. 

CMB secondary anisotropies and polarization exploiting shortly available instrumentations (as for example ALMA) remain the
prime tool to investigate reionization. More along the direction of detecting the ionizing sources are
the number counts of faint objects and  Ly$\alpha$ emitters with NGST, and H$_2$ FIR/sub-mm line emission from PopIII
objects with SIRTF. In the radio, HI 21-cm line tomography of the neutral IGM excited by Ly$\alpha$ pumping should 
be possible with the Square Kilometer Array. High$-z$ SNe (NGST) represent clean probes of 
the cosmic star formation rate and can be used to trace very low mass objects which would be otherwise too faint to be seen.  
Pair production opacity of $\gamma$-ray BLacs can be used to test       the UV background by using EGRET satellite.
Finally, proposals have been made to search for the H$\alpha$ forest {\it if} a H level inversion can be maintained by
some means.

\subsection{A Challenge to Theorists: the TsuCube}           

The ongoing efforts to include radiative transfer into numerical simulations of reionization
deserve a few lines on its own due to their difficulty and beauty. At least four 
groups are attacking the problem by using different numerical schemes: i) Time dependent ray-tracing;
ii) Time-independent ray-tracing; iii) Monte Carlo method and iv) Local Depth Approximation.

At this stage it appears worth to make a comparison between these different approaches. Therefore
the challenge here is to apply the available codes to a commonly defined, well posed radiative transfer problem.
For example, one could predict the ionization patterns produced by a (many) source(s) with a given 
luminosity/spectrum in a given (complex)
density field. This will be defined as the {\bf TsuCube} (in honor of Tsukuba, guest city of the conference),
and it  will allow to compare the codes in term of accuracy and speed of the scheme, so to guide future work. The details of the initial condition must be part of an agreement and discussion among the interested groups
and I have offered myself to act as a coordinator of such comparison. Therefore, interested groups can   
contact me at {\tt ferrara@arcetri.astro.it} to be part of the challenge. 

\section{The Intergalactic Medium}

Naturally, this section complements nicely the previous one and it is partly included into that.
However, there are at least two issues on which is it useful to spend some lines.

\subsection{IGM Metal Enrichment}       
 
One of the results that have become clear recently, is that low mass galaxies (the ones that are
excellent candidates to start the metal enrichment process) lose gas and metals in a differential way.
Whereas typically 90\% of the metals are lost following a blowout event, at most 10\% of the
gas is expelled from the potential well of the parent galaxy. The escape fraction of energy and metals 
decreases with galactic mass, and, when averaged over the entire galaxy population, one finds that at redshifts 
below 3 about 50\% of the metals produced are ejected into the IGM. In spite of this large amount of
heavy elements in the intergalactic reservoir, still unclear remains the nature of the mechanisms
eventually responsible for the spread of such metals. Supernovae might be doing the job provided that
explosions occur at high enough ($z \approx 10$) redshift. This occurrence might explain the apparent
ubiquity of enriched gas suggested by some absorption experiments in the very low column density Ly$\alpha$
forest.

\subsection{IGM Thermal History} 

Recent re-analysis of available absorption spectra of QSOs seems to indicate that the temperature of
the IGM increases up to a maximum at  $z\approx 3$ and that the equation of state becomes close to 
isothermal. Although based on the uncertain assumption of Voigt line profiles, these results can be  
interpreted as due to the extra heating provided to the gas by the late HeII reionization. Indeed several
recent observations imply that HeII reionization is still patchy around $z \approx 3$, thus lending
support to the previous speculation. 
Also, UV background fluctuations would be expected at epochs close to overlap, the 
amplitude of the fluctuations being inversely correlated with the mean flux intensity. Up to now, 
there is no clear evidence of such trend up to $z\approx 5$, the most distant epoch at which the
experiment can be done. This implies that by that redshift hydrogen was already completely ionized, as expected
from the theoretical models summarized above.

\section{Galaxy Formation}    

The hierarchical paradigm providing the theoretical framework to galaxy formation studies
is based on initial conditions (i.e. cosmological models) which seem to be relatively 
settled, at            least judging from the relatively small variety of cosmologies adopted by
the groups. A revolution in the field might be reasonably expected if future experiments like the
measurement of kinetic Sunyaev-Zel'dovich effect in clusters would reveal a non-Gaussian nature
of the initial density field. The rest of the (dark matter) story is rather well established with  
Press-Schechter theory validity being tested down to very low mass scale, and the merging
reasonably well understood from numerical simulations. 

The magic comes in with the baryons, concerning which our knowledge of many physical mechanisms
is either too limited or too complex to be implemented in current models, depending on processes. 
Both semi-analytical models (SAM) and numerical simulations (NS) have to rely on essentially
the same recipes when they boil down to the star formation and feedback processes, dust and 
-- to a lesser extent --  stellar populations. The reward for this unsatisfactory approach is
a wealth of predictions (Tully-Fisher relation, luminosity functions and density, SEDs, disk sizes,
cosmic star formation history and backgrounds, clustering properties, number counts) with which one 
hopes to cure the anemyc physics included. 

Apparently, scalings among different properties of galaxies are the long-searched-for  relations, as today.
For example, an important law as the Tully-Fisher relation connecting the luminosity with the circular velocity, 
can only be partly reproduced: the slope obtained from the simulations agrees with the experimental one 
but an offset is still present which might not be accounted for by changing the stellar feedback prescription
(even in its more modern "kinetic" formulation). In principle the nature of the problem is clear: the CDM 
halos appear to be too concentrated; for the solution new ideas are desperately searched for, unless
this discrepancy is telling us something more fundamental. 
Another unsolved problem of galaxy formation models, at least of  those based on NS, is the failure to
reproduce the angular momentum of disks (similar to the situation with the first sources described above). 
However, new scaling relations are proposed by identifying planes in the 3D space [radius, luminosity, circular
velocity] which only depend on the mass and spin parameter of galaxies.

Chemo-dynamical models seem to have reached an important success in reproducing the color-magnitude
diagrams of dwarf ellipticals from simple quasi-monolithic collapse models. This is more surprising  
as these objects should be expected to be photoevaporated by the metagalactic UVB, of which no account
has been attempted in most simulations of this type. The gas, according to what said above, is ejected 
outside the virial radius from dwarfs, but {\it not} from large ellipticals. However, models with
initial conditions consistent with CDM fluctuations perform slightly well in reproducing 
a variety of physical properties of dElls.

A final note on two more specific topics: dynamics and the starburst-AGN connection. Concerning the
former there is evidence that galaxies in clusters become less massive in time due to stripping/merging.
Also, coupled SAM+Nbody models predict too few S0s in these environment, a problem that might require
a direct S$\rightarrow$S0 transformation. The dynamics of the merging of galaxies with central black holes might  
left traces as the cusps seen in ellipticals.
Starburst still remain poorly understood phenomena and trying to make the connection to the 
AGN activity might come out to be a good strategy to improve our scenarios. It has to be
kept in mind that it is predicted that only $< 10$\% of the cosmic star formation occurs in bursts.
The physics of accretion is rather well established and might help making the connection. The 
most recent advances in this field including the stellar feedback recognize that this addition
tends to increase the effective viscosity of the gas, thus favoring higher (very likely fluctuating)
accretion rates. Finally, the interesting suggestion that differences between Sy2 and Sy1 galaxies 
might be nicely explained by the presence of two radiatively supported, obscuring walls must be
taken seriously and deserves to be fully developed.


%
%
%
\end{document}